# Local Photon and Graviton Mass and its Consequences

M. Tajmar[*]

*ARC Seibersdorf research GmbH, A-2444 Seibersdorf, Austria*

C. J. de Matos[†]

*ESA-HQ, European Space Agency, 8-10 rue Mario Nikis, 75015 Paris, France*

**Abstract**

We show that non-zero masses for a spin-1 graviton (called graviphoton) leads to considerable gravitomagnetic fields around rotating mass densities, which are not observed. The solution to the problem is found by an equivalent graviphoton mass which depends on the local mass density to ensure the principle of equivalence. This solution, derived from Einstein-Proca equations, has important consequences such as a correction term for the Cooper-pair mass anomaly reported by Tate among many others. Similar results were obtained for the photon mass which is then proportional to the charge density in matter. For the case of coherent matter the predicted effects have been experimentally observed by the authors.

---

[*] Head of Space Propulsion, Phone: +43-50550-3142, Fax: +43-50550-3366, E-mail: martin.tajmar@arcs.ac.at
[†] General Studies Officer, Phone: +33-1.53.69.74.98, Fax: +33-1.53.69.76.51, E-mail: clovis.de.matos@esa.int

# 1. Introduction

It is well known that the mass of the photon in vacuum must be non-zero due to Heisenberg's uncertainty principle[1]. There is also an ongoing discussion in the literature if it is possible that the graviton has a non-zero mass as well due to the measurement of the cosmological constant in the universe[2-4]. Both Heisenberg's limit and the cosmological constant lead respectively to a photon and a graviton mass of about $10^{-69}$ kg. This is obviously very small and it is therefore believed that it has negligible consequences. We will show that this is actually not the case and leads to fundamental new insights both for classical and quantum matter. An immediate consequence is that Maxwell equations transform into Proca type equations leading us to the conclusion that for the electromagnetic and gravitational interaction gauge invariance only applies to a certain approximation in free space. The consequence of this result will be discussed in the following. Especially the new treatments required for the graviton have far reaching consequences that can be experimentally assessed.

# 2. Proca Equations and the Photon Mass

As the photon's mass is non-zero, the usual Maxwell equations for electromagnetism transform into the well known Proca equations with additional terms due to the finite Photon wavelength

$$\begin{aligned}
\text{div}\,\vec{E} &= \frac{\rho}{\varepsilon_0} - \left(\frac{m_\gamma c}{\hbar}\right)^2 \cdot \varphi \\
\text{div}\,\vec{B} &= 0 \\
\text{rot}\,\vec{E} &= -\frac{\partial \vec{B}}{\partial t} \\
\text{rot}\,\vec{B} &= \mu_0 \rho \vec{v} + \frac{1}{c^2}\frac{\partial \vec{E}}{\partial t} - \left(\frac{m_\gamma c}{\hbar}\right)^2 \cdot \vec{A}
\end{aligned} \tag{1}$$

The usual effect attributed to a finite Photon mass is its consequence on the strength of electromagnetic forces over distances. However, taking the curl of the 4$^{th}$ equation reveals another feature, which was first assessed by the authors in the framework of superconductivity[5]:

$$B = B_0 \cdot e^{-\frac{x}{\lambda_\gamma}} + 2\omega\rho\mu_0 \lambda_\gamma^2 \;, \tag{2}$$

The first part of Equ. (2) is the Yukawa-type exponential decay of the magnetic field, and the second part shows that a magnetic field will be generated due to the rotation of a charge density $\rho$. In quantum field theory, superconductivity is explained via a large photon mass as a consequence of gauge symmetry breaking and the Higgs mechanism. The photon wavelength is then interpreted as the London penetration depth and leads to a Photon mass about 1/1000 of the electron mass. This then leads to

$$B = B_0 \cdot e^{-\frac{x}{\lambda_L}} - 2\frac{m^*}{e^*}\omega \;, \tag{3}$$

where the first term is called the Meissner-Ochsenfeld effect (shielding of electromagnetic fields entering the superconductor) and the second term is known as the London moment (minus sign is due to the fact that the Cooper-pairs lag behind the rotation) with $m^*$ and $e^*$ as the Cooper-pair's mass and charge (called Becker's argument[16]). The magnetic field produced by a rotating superconductor as a consequence of its large photon mass was experimentally measured outside of the quantum condensate where the photon mass is believed to be close to zero[8,9].

Knowing the effect from a large photon mass in superconductivity, what is the effect of a non-zero photon mass in normal matter? Contrary to the case of superconductivity where the photon mass in comparable to the mass on an electron, in normal matter, the photon mass is believed to be close to zero. The problem of the Proca equations is easily shown: The "Meissner" part becomes important only for large photon masses (which is not the case in normal matter) – but the "London Moment" part becomes important for very small photon masses. By taking the presently accepted experimental limit on the photon mass[1] ($m_\gamma < 10^{-52}$ kg), we can write the second part as

$$B > \omega\rho \cdot 2.8 \times 10^{13} \ . \qquad (4)$$

This would mean that a charge density $\rho$ rotating at an angular velocity $\omega$ should produce huge magnetic fields. Obviously, this is not the case leading to a paradox for normal matter. Therefore the value of the photon mass for matter containing a charge density must be different from the one in free space.

In fact, there is only one possible choice for the photon mass. Let's consider the case of a single electron moving in a magnetic field. It will then perform a precession movement according to Larmor's theorem ω=-(q/2m).B. We can then solve Equ. (2) for the photon mass expressing it as

$$\frac{1}{\lambda_\gamma^2} = \left(\frac{m_\gamma c^2}{\hbar}\right)^2 = -\frac{q}{m} \cdot \mu_0 \rho \ . \tag{5}$$

We find that the photon mass inside normal matter must be defined over the charge density and the charge-to-mass ratio observed. Of course, as the Larmor theorem describes pseudo forces in rotating reference frames, the photon mass in Equ. (5) is not a real mass but can be interpreted as an equivalent photon mass inside the material necessary to comply with Newton's mechanics. Note that this equivalent photon mass is then (due to the negative sign in the Larmor theorem) always a complex value independent of the sign of charge. This is opposite to the case of superconductivity where the photon mass due to the Higgs mechanism indeed has a real value.

Only for neutral matter or vacuum, the photon mass can be therefore given by the limit obtained via Heisenberg's uncertainty principle. As nearly all matter in the universe can be considered neutral, this consequence may be of minor importance. However, in case of gravity and the graviton, the consequences are far reaching as matter is never neutral in a gravitational sense.

## 3. Proca Equations and Graviphoton Mass

In the weak field approximation, gravity can be written similar to a Maxwellian structure forming the so-called Einstein-Maxwell equations. The quantization of the Maxwellian theory of gravity would lead to a spin one boson as a mediator of gravitoelectromagnetic fields. This represents a major problem with respect to the theory of general relativity, which only predicts quadrupolar gravitational waves associated with spin two gravitons. This is the reason why the linear approximation of Einstein field equations is taken as being only an approximation to the complete theory, which cannot be used to investigate radiative processes. Recent experimental results on the gravitomagnetic London moment[6] tend to demonstrate that gravitational dipolar type radiation associated with the Einstein-Maxwell equations is real. This implies that Maxwellian gravity is not only an approximation to the complete theory, but may indeed reveal a new aspect of gravitational phenomena associated with a vectorial spin 1 gravitational boson, which we might call the graviphoton. As an example, a fully relativistic modified theory of gravity called scalar vector tensor gravity[7] would duly take into account this new side of gravity. In the following we will therefore use the term graviphoton for studying the Proca type character of gravity and its consequences on coherent matter.

The field equations for massive linearized gravity are given by[11]:

$$div\,\vec{g} = -\frac{\rho_m}{\varepsilon_g} - \left(\frac{m_g c}{\hbar}\right)^2 \cdot \varphi_g$$

$$div\,\vec{B}_g = 0$$

$$rot\,\vec{g} = -\frac{\partial \vec{B}_g}{\partial t}$$

$$rot\,\vec{B}_g = -\mu_{0g}\rho_m \vec{v} + \frac{1}{c^2}\frac{\partial \vec{g}}{\partial t} - \left(\frac{m_g c}{\hbar}\right)^2 \cdot \vec{A}_g$$

(6)

where $g$ is the gravitoelectric and $B_g$ the gravitomagnetic field. Applying again a curl on the 4th Proca equation leads to

$$B_g = B_{0,g} \cdot e^{-\frac{x}{\lambda_g}} - 2\omega\rho_m \mu_{0g}\lambda_g^2 \,,$$

(7)

where $\rho_m$ is now the mass density and $\mu_{0g}$ the gravitomagnetic permeability (note the different sign with respect to Equ. (2)). Similar to electromagnetism, we obtain a Meissner and a London moment part for the gravitomagnetic field generated by matter[5].

### 4. Consequences of Local Graviton Mass

Applying a non-zero graviphoton mass to Equ. (7) leads again to huge gravitomagnetic fields for rotating mass densities which are not observed. Again we find the solution in the gravitational analog to the Larmor theorem. Locally, the principle of equivalence must be fulfilled for any type of matter. That means that local accelerations must be equivalent to gravitational fields and a body can not distinguish between being in a rotating reference frame or being subjected to a gravitomagnetic field ($B_g = -2\omega$).

By choosing the graviphoton wavelength proportional to the local density of matter similar to Equ. (5),

$$\frac{1}{\lambda_g^2} = \left(\frac{m_g c}{\hbar}\right)^2 = \mu_{0g} \rho_m ,\tag{8}$$

we find the solution. Note that due to the different signs in the Einstein-Proca equations, we find that the equivalent graviphoton mass inside normal matter is a real number to comply with the Larmor theorem in normal matter. Inserting Equ. (8) into our Proca equations for gravity in Equ. (6), we find by performing another grad operator on the 1st equation and another curl on the 4th equation

$$g = -a, \quad B_g = -2\omega ,\tag{9}$$

which is nothing else as the formulation of the equivalence principle and the gravitomagnetic Larmor theorem[10]. The equivalent graviphoton mass in Equ. (8) therefore describes the inertial properties of matter in accelerated reference frames. This is a very fundamental result and new insight into the foundations of mechanics.

A similar graviphoton mass (up to a factor 3/16) was already found by Argyris[11] by solving Einstein's equations in the conformally flat case. He linked it with the average mass of the universe. However, as we have shown, this result is valid locally for all matter.

It is interesting to note that if we take the case of no local sources ($\rho_m = 0$), the graviphoton mass will be zero, and we will find, by solving the weak field equations in the transverse gauge, the "classical" freely propagating degrees of freedom of gravitational waves associated with a massless spin 2 graviton[12]. However, in the case of local sources, a spin-1 graviphoton will appear.

**4.1 Application to Coherent Quantum Matter**

Perhaps that most important consequence of the local graviphoton mass is its relation to superconductivity. As we wrote already in the introduction of this paper, the application of Proca equations to superconductivity are well established. In a superconductor, we have now a ratio between matter being in normal and in a condensed (coherent) state. So we have two sets of Proca equations, one which deals with the overall mass and one with its condensated subset.

The coherent part of a given material (e.g. the Cooper-pair fluid) is also described by its own set of Proca equations similar to the ones in Equ. (6) but with one important difference: Instead of the ordinary mass density $\rho_m$ we have to take the Cooper-pair mass density $\rho_m^*$. By taking the curl of the fourth Proca equation (See Ref 5), we arrive at

$$B_g = B_{0,g} \cdot e^{-\frac{x}{\lambda_g}} + 2\omega \rho_m^* \mu_{0g} \lambda_g^2 \ , \tag{10}$$

where we had to introduce Becker's argument[16] that the Cooper-pairs are lagging behind the lattice so that the current is flowing in the opposite direction of $\omega$. This gives the right sign with respect to our experimental observation[6]. A similar argument was introduced for the classical London moment as also here the sign change was observed accordingly. The first part (Meissner part) is not different from our previous assessment for normal matter, but the second part changes to

$$B_g = 2\omega \frac{\rho_m^*}{\rho_m} \ , \tag{11}$$

due to the fact that the graviphoton mass depends on all matter in the material, not just the coherent part. That is why the densities do not cut any more and we arrive at an additional field for coherent matter in addition to the gravitational Larmor theorem. As an alternative to Becker's arguments[16] on the sign change in superconductors, one could also switch between real and imaginary values for the equivalent photon and graviphoton when going from the normal to the coherent state of matter.

Similarly, by taking the gradient of the first Proca equation in Equs (6), considering the case of an homogeneous gravitational field, and using the fact that the gradient of a density of energy is equal to a density of force, $\nabla(\rho_m^* c^2) = \rho_m^* \vec{a}$, we obtain:

$$\vec{g} = \vec{a} \mu_{0g} \rho_m^* \lambda_g^2 \ , \tag{12}$$

which transforms using Equ.(9), into:

$$\vec{g} = -\frac{\rho_m^*}{\rho_m} \vec{a} \ , \tag{13}$$

where $\vec{a}$ is the total net acceleration to which the superconductor is submitted.

Comparing Equs (11) and (13) to Equs (9) we conclude that the presence of Cooper-pairs inside the superconductor leads to a deviation from the equivalence principle and from the classical gravitational Larmor theorem. A rigid reference frame mixed with non-coherent and coherent matter is not equivalent to a rigid reference frame made of normal matter alone, with respect to its inertial and gravitational properties. However, in the case of a Bose-Einstein condensate where we have only coherent matter, Equ. (13) transforms into the usual expression for normal matter and the equivalence principle is again conserved.

An important feature is that these fields (Equs. (11) and (13)) however, should be also present <u>outside</u> the superconductor, contrary to the classical inertial behaviour. The gravitomagnetic Larmor theorem for normal matter describes the inertial forces in an accelerated reference frame. This leads to so-called pseudo forces which are only present inside the material which is rotating. The difference to quantum materials is that in this case, the integral of the canonical momentum is quantised. Let's consider a superconducting ring. The integral of the full canonical momentum of the Cooper-pairs including gravitational fields is given by

$$\oint \vec{p}_s \cdot d\vec{l} = \oint \left( m^* \vec{v}_s + e^* \vec{A} + m^* \vec{A}_g \right) \cdot d\vec{l} = \frac{nh}{2} \; . \tag{14}$$

If the ring is thicker than the London penetration depth, then the integral can be set to zero. Solving for the case where the superconductor is at an angular velocity $\omega$, we get

$$\vec{B} = -\frac{2m}{e} \cdot \vec{\omega} - \frac{m}{e} \cdot \vec{B}_g . \qquad (15)$$

These magnetic and gravitomagnetic fields are also present inside the superconducting ring. The first part is the classical London moment, with its origin is due to the photon mass, and the second part is its analog gravitomagnetic London moment, which will produce an additional field overlapping the classical London moment. According to Equ. (11), depending on the superconductor's bulk and Cooper-pair density, the magnetic field should be higher than classically expected. Indeed, that has been measured without apparent solution throughout the literature.

The authors already conjectured such a field to explain a reported disagreement between the theoretical and experimental Cooper-pair mass[5,17,18]. Tate et al[8,9] used a sensitive London moment measurement to determine the Cooper-pair mass in Niobium. This mass was found to be larger ($m^*/2m_e$ = 1.000084(21)) than the theoretically expected value ($m^*/2m_e$ = 0.999992). As we pointed out earlier in our conjecture, this mass difference opens up the room for large gravitomagnetic fields following the quantised canonical momentum in Equ. (12). In order to correct Tate's result, we need a gravitomagnetic field of

$$B = 2 \frac{m^*_{experimental} - m^*_{theoretical}}{m^*_{theo}} \omega = 2 \frac{\Delta m^*}{m^*} \omega , \qquad (16)$$

where $\Delta m^*$ is the difference between the experimental and theoretical Cooper-pair mass to find back the Cooper-pair mass predicted by quantum theory.

The local graviton mass now establishes the reason why such a gravitomagnetic field has to be there. Comparing Equs. (11) and (16), we can identify

$$\frac{\Delta m^*}{m^*} = \frac{\rho_m^*}{\rho_m} \quad . \tag{17}$$

Taking Tate's values ($\frac{\Delta m^*}{m^*} = 9.2 \times 10^{-5}$) and the Niobium bulk and Cooper-pair mass density ($\frac{\rho_m^*}{\rho_m} = 3.95 \times 10^{-6}$), we see that these values are a factor of 23 away. One has to take into account that Tate's measurement is up to now the only precision experiment and, even more important, other relativistic correction terms need to be added to the theoretical Cooper-pair mass, which will make *Δm\** smaller and Equ. (17) match better.

This is a very important result as we have for the first time not only a conjecture to explain Tate's anomaly, but also a good reason why a rotating superconductor should produce a gravitomagnetic field which is larger than classical predictions from ordinary rotating matter. The reason is the local (equivalent) graviphoton mass. **Fig. 1** plots the predicted Δm/m for various superconductors.

In parallel, the authors also attempted to measure the gravitomagnetic field from induced gravitational fields around a rotating superconductor which is reported in Ref. 6. First measurements show that this gravitomagnetic field indeed exists with a measurement in between our Cooper-pair density ratio and the one derived from

Tate's measurements ($\frac{\Delta m^*}{m^*}_{measured} \cong 2.6 \times 10^{-5}$). This adds strong confidence in our theoretical approach and its consequences.

**Conclusion**

We have shown that non-zero values for the graviphoton leads to huge gravitomagnetic fields around rotating mass densities, which are not observed. The solution to the problem is found by an equivalent graviphoton mass which depends on the local mass density leading to the correct inertial forces in rotating reference frames. That can be understood as a foundation of basic mechanics. This solution, derived from Einstein and Proca equations, has important consequences such as

- for the case of normal matter, the Proca equations now lead to the gravitomagnetic Larmor theorem and not to unobserved huge gravitomagnetic fields,
- the prediction of a gravitomagnetic London moment (observed experimentally) in rotating superconductors, that can solve the Cooper-pair mass anomaly reported by Tate, among many others. Similar results have also been outlined for the Photon mass. By obeying Larmor's theorem, we find that in classical matter the equivalent photon has a complex and the graviphoton a real value. In coherent matter we suggest the hypothesis that it is exactly the other way round, which solves the sign change problems associated to the classical and gravitomagnetic London moment as an alternative to the usual Becker argument. The results are very encouraging and shall stimulate the further development of the basic concept outlined in this paper:

## References

[1] Tu, L., Luo, J., Gillies, J.T., *Rep. Prog. Phys.* **68**, 77-130 (2005).

[2] Spergel, D.N., et al., *Astrophy. J. Suppl.*, **148**, 175 (2003).

[3] Novello, M., Neves, R.P., *Class. Quantum Grav.* **20**, L67-L73 (2003).

[4] Liao, L., gr-qc/0411122, 2004

[5] De Matos, C.J., Tajmar, M., *Physica C* **432**, 167-172 (2005).

[6] Tajmar, M., Plesescu, F., Marhold, K., de Matos, C.J., *Physica C (submitted)*

[7] Moffat, J.W., gr-qc//0506021

[8] Tate, J., Cabrera, B., Felch, S.B., Anderson, J.T., *Phys. Rev. Lett.* **62**(8), 845-848 (1989).

[9] Tate, J., Cabrera, B., Felch, S.B., Anderson, J.T., *Phys. Rev. B* **42**(13), 7885-7893 (1990).

[10] Mashhoon, B., *Phys. Lett. A*, **173**, 347-354 (1993).

[11] Argyris, J., Ciubotariu, C., *Aust. J. Phys.*, **50**, 879-891 (1997).

[12]Carroll, S., Spacetime and Geometry: An Introduction to General Relativity ( Addison Wesley; 1st edition, 2003), 293-300

[13]Neupane, I.P., gr-qc/9812096

[14]Obukhov, Y.N., Vlachynsky, E.J., *Ann. Phys.* **8**(6), 497 – 509 (1999).

[15]Bender, R., *The Astrophysical Journal* **631**, 280-300 (2005).

[16]Becker, R., Heller, G., Sauter, F., *Z. Physik* **85**, 772-787 (1933).

[17]Tajmar, M., de Matos, C.J., Gravitomagnetic Field of a Rotating Superconductor and of a Rotating Superfluid", *Physica C*, **385**(4), 2003, pp. 551-554

[18]Tajmar, M., de Matos, C.J., *Physica C* **420**(1-2), 56-60 (2005).

[19]Rolnick, W.B., Fundamental Particles and their Interactions (Addison-Wesley Publishing Company, 1994)

[20]Renton, P., *Nature*, **428**, 141 (2004).

[21]Delgado, V., hep-ph/9305249

[22]Delgado, V., hep-ph/9403247

| Location | Cosmological Constant [m²] |
|---|---|
| Sun | $1.97 \times 10^{-23}$ |
| Earth | $7.68 \times 10^{-23}$ |
| Solar System | $3.14 \times 10^{-35}$ |
| Milky Way | $6.29 \times 10^{-48}$ |
| Universe | $1.29 \times 10^{-52}$ |

**Table 1**  Cosmological Constant Examples

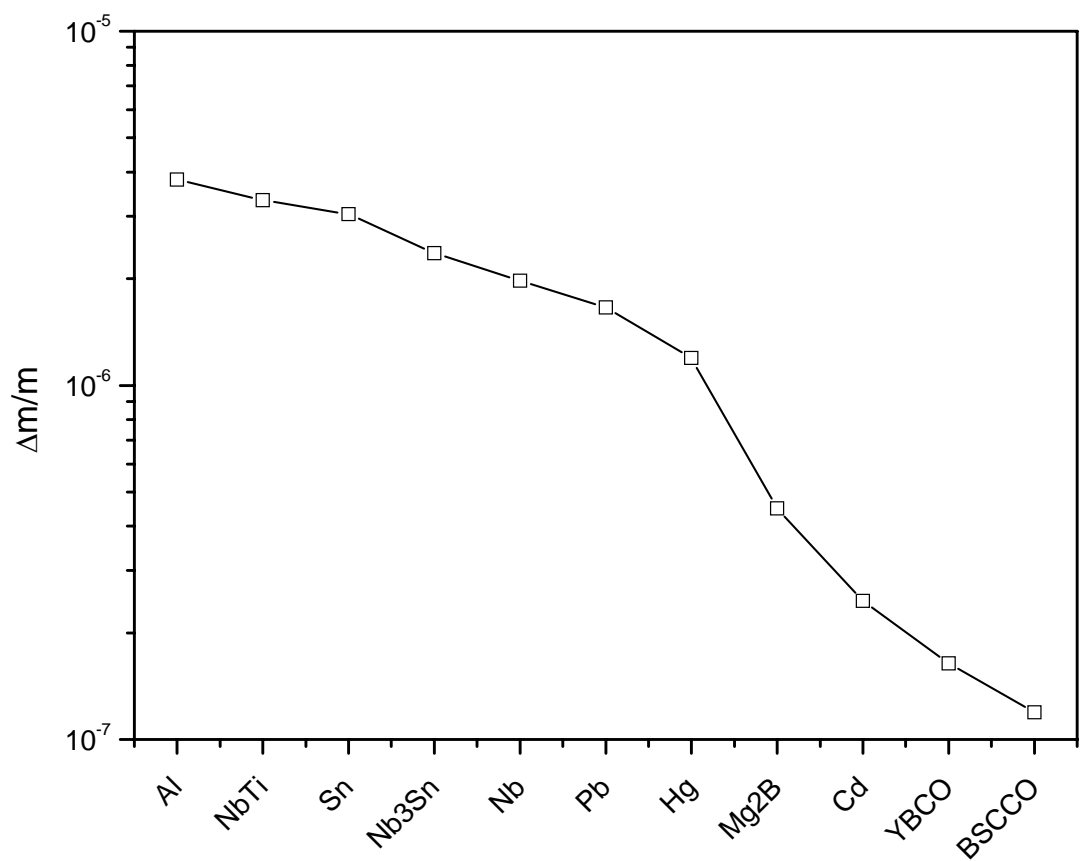

**Fig. 1** Predicted Δm/m for various Superconductors